\documentclass[letterpaper,twocolumn,showpacs,preprintnumbers,amsmath,amssymb]{revtex4}

\usepackage{graphicx}
\usepackage{dcolumn}
\usepackage{bm}

\DeclareGraphicsExtensions{.eps,.EPS}

\begin{document}

\preprint{Submitted for publication.}

\title{Determination of the $^{85}$Rb 
$ng$-series quantum defect by electric-field-induced
resonant energy transfer 
between cold Rydberg atoms
}%

\author{K.~Afrousheh}
\author{P.~Bohlouli-Zanjani}
\author{J.~A.~Petrus}
\author{J.~D.~D.~Martin}
\affiliation{%
Department of Physics and Astronomy and 
Institute for Quantum Computing \\
University of Waterloo, Waterloo, Ontario, N2L 3G1, Canada
}%

\date{\today}

\begin{abstract}

Resonant energy transfer between cold Rydberg atoms was
used to determine Rydberg atom energy levels, at precisions
approaching those obtainable in microwave spectroscopy.
Laser cooled $^{85}{\rm Rb}$ atoms from a 
magneto-optical trap were optically excited to $32d_{5/2}$ 
Rydberg states.  The two-atom process 
$32d_{5/2}$ + $32d_{5/2} \rightarrow 34p_{3/2} + 30g$
is resonant at an electric field 
of approximately $0.3 \: {\rm V/cm}$.  This process is driven
by the electric dipole-dipole interaction, which is allowed due to
the partial $f$ character that the $g$ state acquires in
an electric field.  The experimentally observed resonant field,
together with the Stark map calculation is used to make
a determination of the $^{85}$Rb $ng$-series
quantum defect: $\delta_{g} (n=30) = 0.00405(6)$.

\end{abstract}


\maketitle

\section{Introduction}

Energy can be transferred more rapidly between two atoms when
the process is {\em resonant}: one atom gives its energy to
another, without a change in the kinetic energy of either atom.
Although resonant energy transfer often relies on serendipity,
the large Stark shifts of Rydberg atoms may allow the
resonant condition to be exactly obtained by application
of an electric field \cite{safinya:1981}. The resonant energy transfer
process between Rydberg atoms is driven by the electric 
dipole-dipole interaction.  Transition dipole moments
between nearby Rydberg states can be large, typically scaling as
$n^2$, where $n$ is the principal quantum number \cite{gallagher:1994}.
As a consequence,
resonant energy transfer collision cross-sections scale as $n^4$,  
and are large compared to typical collision cross-sections for less 
excited atoms -- usually by many orders of magnitude 
\cite{safinya:1981,gallagher:1994}.

Stoneman {\it et al.}~\cite{stoneman:1987} pointed out
that the fields at which resonant energy transfer occurs may
-- under certain conditions --
be used for precise Rydberg energy level spectroscopy.
They examined the resonant
energy transfer process: $29s + 29d \rightarrow 29p + 28p$,
in potassium.  Since the shifts of the Rydberg energy levels
with field are relatively well-understood, they were able
to use the resonance fields to determine improved zero field
atomic energy levels.  In particular, they determined a
substantially more precise value for the quantum defect of 
the $np$-series of potassium.

The range of electric fields or ``width'' over which significant 
resonant energy  transfer can be observed is important
in this type of experiment, as it dictates the precision to
which line centers can be determined.  
By reducing the spread in the velocity distribution of their sample
Stoneman {\it et al.}~\cite{stoneman:1987} 
were able to significantly narrow the width
of their observed resonances.

With the development of laser cooling,
Rydberg atom samples with much narrower velocity distributions 
could be studied \cite{anderson:1998,mourachko:1998}.
In general, the resulting narrowness of the resonant energy
transfer resonances has not been exploited for spectroscopic purposes.
Instead, work has concentrated on the dynamical aspects of resonant 
energy transfer between cold Rydberg atoms. For example,
Carroll {\it et al.}~\cite{carroll:2004}
have recently studied how the relative orientation of the electric field
with respect to the spatial orientation of two Rydberg atoms influences
their dipole-dipole interaction.  They used cold Rb
atoms obtained from a magneto-optical trap (MOT) to study the process:
$32d + 32d \rightarrow 34p + 30k$,
where $30k$ represents the Stark-states centered around $n = 30$.
The Stark states are superpositions of high angular momentum 
states ($\ell \gtrsim 4$ in this case) 
exhibiting energy shifts which are linear in the applied 
electric field strength.
Strong resonant energy transfer was observed between 
$1$ to $6 \: {\rm V/cm}$,
and was fully explained by a Stark map calculation, which
took into account the energy shifts of the participating atoms.
However, a series of peaks at approximately $0.5 \: {\rm V/cm}$
were not reproduced in the calculations
(these peaks were not relevant to the main point of their paper).

In this paper we demonstrate that the resonances
observed at low field are due to the process:
\begin{equation}
\label{eq:us}
32d + 32d \rightarrow 34p_{3/2} + 30g
\end{equation}
which is an allowed electric dipole-dipole interaction, 
due to the $f$ character that the $g$ state
acquires in the electric field.  
In the spirit of the work of Stoneman {\it et al.}~\cite{stoneman:1987} 
we use the observed resonant fields together with a Stark map calculation
to determine the $ng$-series quantum defect.  

Han {\it et al.}~\cite{han:2006} have reported a preliminary observation of
two-photon $(n+2)d-ng$ microwave transitions in $^{85}$Rb.
However, we are unaware of any other work on the $ng$ Rydberg series of Rb.  

\section{Experiment}

The essence of the experiment is to measure the electric
fields required for efficient resonant energy transfer between
cold $32d_{5/2}$ $^{85}$Rb atoms.
Some details of our apparatus have been previously
described \cite{afrousheh:2006}.  
A standard vapor cell MOT is used as the source of cold $^{85}$Rb atoms.
These atoms are excited to 32$d_{5/2}$ Rydberg states using a 
two-photon, two-color process with 480 nm light, and the
nearly resonant, red-detuned 780 nm light used for cooling
and trapping.  The 480 nm light is obtained by frequency doubling
a 960 nm cw ring Ti:sapphire laser, which is frequency stabilized using
the technique described in Ref.~\cite{bohlouli:2006}.
The experiment is done at a $10 \: {\rm Hz}$ repetition rate,
with an acousto-optic modulator used to produce pulses of light.

For energy level determination from resonant energy transfer
spectroscopy it is critical to zero out any stray electric fields,
and then be able to apply accurately known electric fields.
Two stainless steel electrode plates, separated
by $36 \: {\rm mm}$, are located to either side of the trapped atoms.
By varying the voltages on these plates and the voltage 
of the Rb dispenser source with
respect to the grounded chamber, it is possible to compensate
the stray electric field in all three directions.
The single-photon transition $48s_{1/2}-48p_{1/2}$ in
$^{85}$Rb is used for this purpose.  
This general procedure has been described by
Osterwalder and Merkt \cite{osterwalder:1999}.
Since the single photon transition is sensitive to magnetic fields,
the inhomogeneous magnetic field necessary for MOT operation is
shut-off prior to photoexcitation to Rydberg states.  At
the time of the experiment the residual magnetic field 
inhomogeneity over the sample is less than $17 \: {\rm mG}$.
The details of this shutting off procedure
are presented in Ref.~\cite{afrousheh:2006}.

Once the stray electric field is compensated, the plate voltages
may be used to produce deliberate fields.  We calibrate these
fields by observing the shifted frequencies of the
$48s_{1/2}-48p_{1/2}$ transition and comparing them to the
frequencies expected from a Stark map calculation \cite{zimmerman:1979}.
The Stark map calculation uses the most recent quantum defect data
for $^{85}$Rb \cite{li:2003,han:2006}.  Although quantum defects
for $\ell > 3$ are not known, these do not significantly
influence the calculated transition frequencies.

To observe resonant energy transfer between Rydberg atoms with
known electric fields, we produce the fields
in exactly the same way as during calibration using
the $48s_{1/2}-48p_{1/2}$ transition.  The inhomogeneous magnetic
field is also shut-off in an identical manner.
After a light excitation pulse of $1 \: {\rm \mu s}$ duration,
the electric field is ramped from zero to the desired value in
$160 \: {\rm ns}$.  The atoms may interact during
a waiting period of $21 \: {\rm \mu s}$.

\begin{figure}
\includegraphics{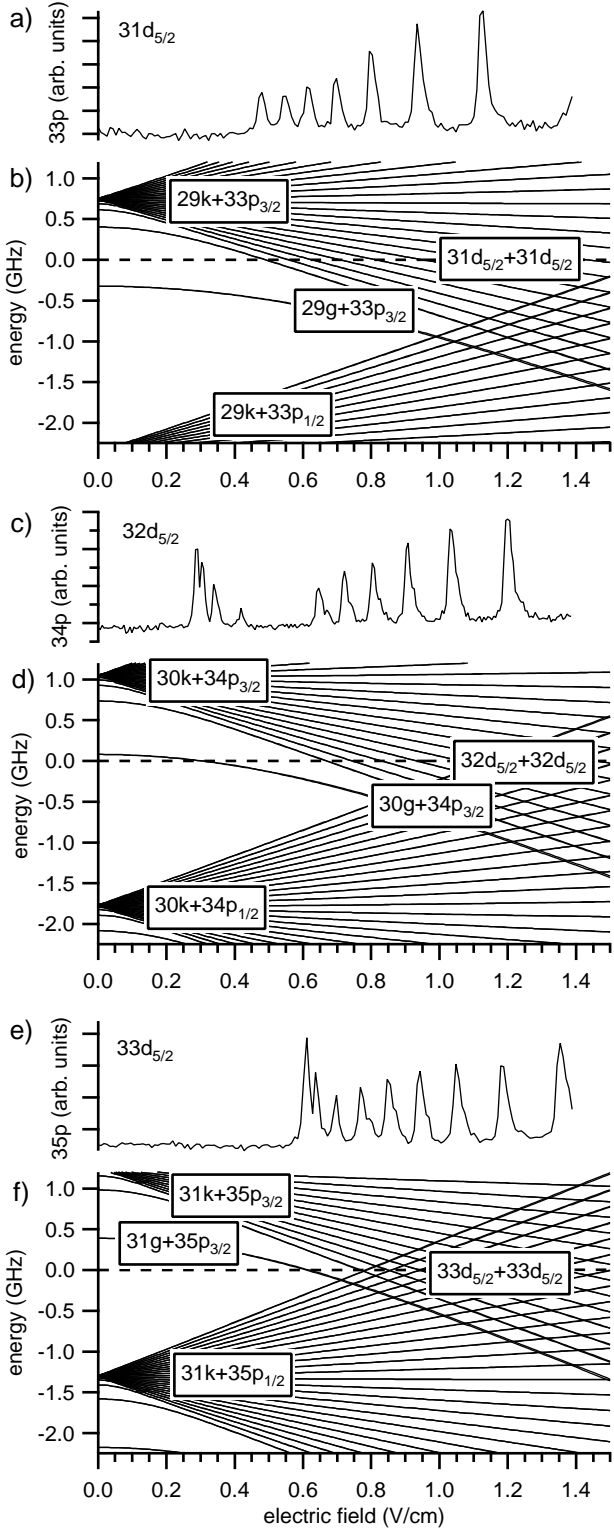}
\caption{\label{fg:spectra}
(a,c,e) Resonant energy transfer spectra with atoms initially in
the $31d_{5/2}$, $32d_{5/2}$ and $33d_{5/2}$ states.
(b,d,f) Calculated total energy of several two atom states, relative
to the zero field energy of $nd_{5/2}+nd_{5/2}$ where 
$n=31,32,33$ for (b), (d) and (f) respectively.  The calculations
follow the procedures of Zimmerman {\it et al.}~\cite{zimmerman:1979},
using $\delta_g=0.004$ and for $\ell >4$,
$\delta_\ell = \delta_g (4/\ell)^5$. 
}
\end{figure}

Final Rydberg state populations after the waiting period are
analyzed by selective field ionization (SFI) \cite{gallagher:1994}.  
In this case SFI is only able to distinguish states differing by 
energies equivalent to a change of approximately
one principal quantum number.  
The 34$p$ Rydberg states are ionized at a lower field than
both the $32d_{5/2}$ initial states and 30$g,k$ final states.
By changing the applied electric field and
recording the 34$p$ population, resonant energy transfer spectra
may be obtained (see Fig.~\ref{fg:spectra}c).  
For comparison
we have also recorded analogous spectra at one lower $n$
(Fig.~\ref{fg:spectra}a) 
and one higher $n$ (Fig.~\ref{fg:spectra}e).

\section{Analysis}

The striking difference in the three spectra presented in 
Fig.~\ref{fg:spectra}
is the presence of the low-field resonances in the $32d_{5/2}$
case, but not in the others.  Carroll 
{\it et al.}~have also observed similar resonances
in $32d$ (Ref.~\cite{carroll:2004}), but not for $31d$ 
(Ref.~\cite{carroll2:2006}).  
Energetics \cite{li:2003,han:2006} rule out the process:
$32d_{5/2}+32d_{5/2} \rightarrow 34p_{3/2} + 30f$ as a possible
explanation for these low-field resonances.
The final state $34p_{3/2} + 30f$ 
is at an energy less than the initial state in zero electric field,
and this difference increases with increasing field,
due to the $f$ state polarizability.  

There are no processes involving the well-characterized
low-angular momentum Rydberg states  ($\ell \le 3$) of Rb
which would account for the low field resonances in Fig.~\ref{fg:spectra}c.
We need to consider higher angular momentum states.
The non-penetrating high angular momentum states of Rydberg atoms
have quantum defects which scale like $1/\ell^5$
(see, for example, Ref.~\cite{gallagher:1994}).
Thus a rough estimate of the $ng$-series quantum defect
may be obtained from the $nf$-series quantum defect \cite{han:2006}: 
$\delta_g \approx \delta_f (3/4)^5 \approx 0.004$.
A preliminary calculation with this estimated
quantum defect shows that a resonance will occur for the process:
$32d_{5/2}(m_j\!=\!1/2) + 32d_{5/2}(m_j\!=\!1/2) \rightarrow 
34p_{3/2}(m_j\!=\!1/2) + 30g_{7/2}(m_j\!=\!1/2)$
at an electric field of $0.31 \: {\rm V/cm}$.
This is very close to the lowest field resonance observed in
Fig.~\ref{fg:spectra}c.  However for the $31d_{5/2}$ case,
the energy of the two-atom state $33p_{3/2}+29g$
is less than the initial state $31d_{5/2}+31d_{5/2}$, and
this difference increases with increasing field 
(see Fig.~\ref{fg:spectra}b). 
This is consistent with the lack of observed low-field resonances
in the $31d_{5/2}$ spectrum (Fig.~\ref{fg:spectra}a).

In the case of the $33d_{5/2}$, the electric field
does shift $35p_{3/2}+31g$ closer in energy to $33d_{5/2}+33d_{5/2}$.
However, the $31g$ state reaches the $n=31$
Stark manifold at approximately the same field
(Fig.~\ref{fg:spectra}f).
This is consistent with the experimental observations in 
Fig.~\ref{fg:spectra}e.

The calculations in Fig.~\ref{fg:spectra} 
consider only $m_j=1/2$ magnetic sublevels.  
However, electric fields break the magnetic sublevel 
energy degeneracy; 
therefore, there should be different resonance
fields corresponding to the different possible magnetic sublevels
for the states in Eq.~\ref{eq:us}.  The largest splittings are
due to the magnetic sublevels of $30g$.
The four peaks observed at low fields
in Fig.~\ref{fg:spectra}c
correspond to $m_{\ell}=$ 0, 1, 2, 3, and 4 (the $m_{\ell}=$ 0
and 1 resonances are unresolved).
The splittings due to the magnetic sublevels of the other
states are much less than the widths of each of these four peaks.
Similarly, the three peaks observed between 0.6 and 
$0.7 \: {\rm V/cm}$ in Fig.~\ref{fg:spectra}e correspond to
the magnetic sublevels of $31g$.

\begin{figure}
\includegraphics{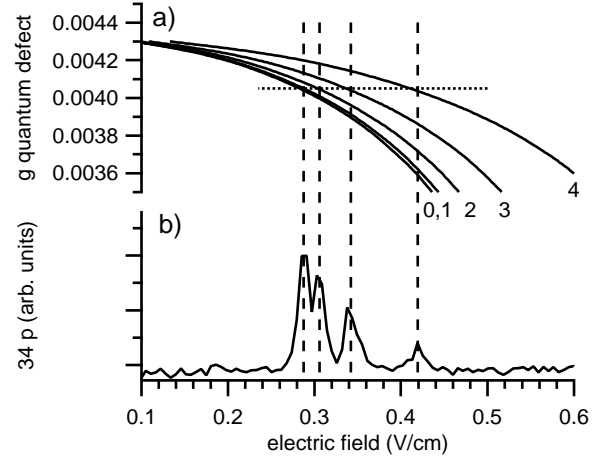}
\caption{\label{fg:fine}
a) Resonance fields calculated for variable $\delta_g$ (vertical axis).
The five lines correspond to 
different magnetic sub-levels ($m_{\ell}$) for the $30g$ state.
The calculated fields are shown on the horizontal axis to allow
a direct comparison to the experimental data below.
b) The same $32d_{5/2}+32d_{5/2}$ resonant energy transfer
spectrum as Fig.~\ref{fg:spectra}c), shown on an expanded
scale.
}
\end{figure}

All four observable resonances should be explainable by the same
$ng$-series quantum defect ($\delta_g$).  
Figure \ref{fg:fine}a illustrates calculated
resonance fields as a function of a variable $\delta_g$,
centered around the estimated $\delta_g \approx 0.004$ (see above).
As illustrated in Fig.~\ref{fg:fine}, all four resonances are consistent
with the same quantum defect, to better than the widths of the 
resonances.  We consider this as strong evidence
that the process in Eq.~\ref{eq:us} is the explanation for the low
field resonances.

As Fig.~\ref{fg:fine} suggests, we can extract an estimate of 
the quantum defect from each one of these four resonances.
The uncertainty may also be estimated for each of
these experimentally determined quantum defects from
the derivatives $\partial \delta_g / \partial F$ evaluated
at the estimated quantum defects (where $F$ is the electric
field strength).
The electric field zeroing and calibration procedure is estimated
to have a $1\sigma$ uncertainty of less than $30 \: {\rm mV/cm}$ 
at the resonance fields.  
From the four resonances we determine
$\delta_{g} (n=30) = 0.00405(6)$, which is consistent
with a preliminary
report of 
$\delta_{g} = 0.00400(9)$
from Han {\it et al.}~\cite{han:2006}.

As demonstrated, energetics strongly support 
Eq.~\ref{eq:us} as the explanation for the low-field resonances.
However, what drives this process?
Resonant energy transfer between Rydberg atoms is
typically a dipole-dipole interaction, with the coupling
between the two-atom states given by:
\begin{equation}
\hat{V}_{dd} = \frac{\vec{\mu}_{\rm A} \cdot \vec{\mu}_{\rm B}
    - 3 (\vec{\mu}_{\rm A} \cdot \vec{n}) 
(\vec{\mu}_{\rm B} \cdot \vec{n})}{R_{\rm AB}^3}
\end{equation}
where $R_{\rm AB}$ is the separation of the two atoms,
$\vec{n}$ is a unit vector pointing between them, and
$\vec{\mu}_{\rm A}$ and $\vec{\mu}_{\rm B}$ are 
{\em transition} dipole moments evaluated on each of the
individual atoms (A and B).  In zero electric
field the process given in Eq.~\ref{eq:us} 
cannot be driven by the dipole-dipole interaction, since
$\left\langle 32d_{5/2}|\vec{\mu}|30g_{7/2} \right\rangle = 0$, given the
$\Delta \ell = \pm 1$ selection rule for transition dipole
moments.  However, when a weak electric field is applied, the
$\ell$ quantum number is no longer good. Instead, it serves
to label the dominant character of a state until it reaches
the Stark manfold.  
In particular, the mixing of $30g$ with $30f$ gives 
$\left\langle
32d_{5/2}m_j\!\!=\!\!1/2|\mu_z|30g_{7/2}m_j\!\!=\!\!1/2\right\rangle 
= 17 \: q_e a_0 $ 
at $0.3 \: {\rm V/cm}$, where $q_e a_0$ is the 
atomic unit of electric dipole moment. 
The $32d_{5/2}$ to $34p_{3/2}$ coupling is
allowed in zero field, and is much stronger.
For example at $0.3 \: {\rm V/cm}$, 
$\left\langle
32d_{5/2}m_j\!\!=\!\!1/2|\mu_z|34p_{3/2}m_j\!\!=\!\!1/2\right\rangle= 357
\: q_e a_0$ (which differs little from its zero field value).

It is interesting to compare the expected signal strengths
for the resonance in Eq.~\ref{eq:us}
with other resonances that have been studied in similar 
experimental set-ups.
For example, if we compare our 
transition dipole moments to those for the $^{85}$Rb, 
$33s_{1/2}+25s_{1/2} \rightarrow 24p_{1/2}+34p_{3/2}$
process observed in Ref.~\cite{anderson:1998}, 
we would expect to have to wait approximately 10 times
longer to see the equivalent amount of collision signal (at
the same density).
Since typical waiting times are 
$3 \: {\rm \mu s}$ in Ref.~\cite{anderson:1998}
{\it vs.}~$21 \: {\rm \mu s}$ 
in our case, this is consistent.

In the quasi-static picture \cite{anderson:1998} 
the collision signal is expected to scale 
with the square of the density for a dipole-dipole interaction, 
and to a higher power for higher-order multipole processes
(for small collision signals).
Experimentally, we observe square law behavior.  It does not appear
necessary to consider higher-order multipole interactions
(see, for example, Ref.~\cite{gallagher:1994}).

\section{Concluding Remarks}

Resonant energy transfer spectroscopy has been demonstrated as
a useful tool for the determination of unknown energy levels
differences.
In particular, prior to the work of Han {\it et al.}~\cite{han:2006},
no spectroscopy of the Rb $ng$-series had been performed.  

Although resonant energy transfer spectroscopy is not
a general technique for the determination of energy levels, there may be 
more possibilities for the observation of resonances than previously 
thought.  In particular, at first glance the resonance studied in this paper
(Eq.~\ref{eq:us}) does not appear to be allowed by the dipole-dipole
interaction.  However, the mixing of the $g$-state with the 
nearby $f$-state allows this process to be observed.
It is expected that similar scenarios exist in Rb and other atoms.

Microwave transitions have been used to calibrate the applied electric
fields in this work.
However, there is a great deal of flexiblity in the particular 
transition (and corresponding microwave frequency) that can be used for this 
calibration.  A precise microwave source is not required
for the actual spectroscopic measurement.  
This may be considered to be a significant advantage.

\section{Acknowledgments}

We thank the authors of Ref.~\cite{han:2006} for providing us with
a preprint prior to publication and acknowledge useful
conversations with T. F. Gallagher, M. Noel and J. D. Carter.  
We thank D. Vagale for assistance with the Stark map calculations.
This work was supported by NSERC, CFI, and OIT.

\end{document}